\documentclass[aps,pra,twocolumn,groupedaddress,amsmath,amssymb,amsfonts]{revtex4-1}

\usepackage{mathtools}
\usepackage{subdepth}
\usepackage{glossaries}
\usepackage{hyperref}
\usepackage{xparse}

\let\phi\varphi

\DeclareMathAlphabet\mathbfcal{OMS}{cmsy}{b}{n}

\DeclareMathOperator\PT{\mathcal{PT}}

\DeclareMathOperator\rpart{Re}
\DeclareMathOperator\ipart{Im}

\DeclareMathOperator\Hamiltonian{\mathcal{H}}
\DeclareMathOperator\tr{\mathrm{tr}}
\DeclareMathOperator\var{\mathrm{var}}
\DeclareMathOperator\cov{\Delta}

\DeclarePairedDelimiter\parens\lparen\rparen
\DeclarePairedDelimiter\brackets\lbrack\rbrack
\DeclarePairedDelimiter\angles\langle\rangle
\DeclarePairedDelimiter\abs\lvert\rvert

\DeclarePairedDelimiterX\commutator[2][]{#1\,,\,\mathopen{}#2}
\DeclarePairedDelimiter\expval\langle\rangle
\DeclarePairedDelimiter\bra\langle\rvert
\DeclarePairedDelimiter\ket\lvert\rangle

\newcommand\comma{\nolinebreak\,,}
\newcommand\period{\nolinebreak\,.}
\newcommand\operator[1]{\hat{#1}}

\newcommand\imag{\mathrm{i}\mkern1mu} % see ISO 80000-2:2009
\newcommand\pdiff[3][\empty]{\frac{\partial^{#1}#2}{\partial#3^{#1}}}
\newcommand\absquare[1]{\abs*{#1}^2}

\newcommand\conjg[1]{{#1}^\ast}
\newcommand\adjg[1]{{#1}^\dagger}

\newcommand\BH{\mathrm{BH}}
\newcommand\ad{\adjg{\operator{a}}}
\newcommand\n{\operator{n}}

\newcommand\xpdm[1]{\sigma^{(#1)}}
\newcommand\oxpdm[1]{\operator{\sigma}^{(#1)}}
\newcommand\spdm{\xpdm{1}}
\newcommand\ospdm{\oxpdm{1}}
\newcommand\tpdm{\xpdm{2}}

\newcommand\jt{\skew{3}\tilde{j}}
\newcommand\ojt{\skew{3}\operator{\skew{3}\tilde{j}}}
\newcommand\jtx[1]{\jt^{(#1)}}
\newcommand\cx[1]{c^{(#1)}}

\newcommand\Zeta[1]{\mathcal{Z}^{(#1)}}
\newcommand\Chi[1]{\mathcal{X}^{(#1)}}
\newcommand\Ups[1]{\mathcal{Y}^{(#1)}}

\renewcommand\a{\operator{a}}

\newcommand\subref[2]{\ref{#1} (#2)}
\newcommand\ac[1]{\gls*{#1}}
\newcommand\acp[1]{\glspl*{#1}}

\newcommand\acrodef[3][\empty]{%
  \ifx{#1}\empty%
    \newacronym{#2}{#2}{#3}%
  \else%
    \newacronym[#1]{#2}{#2}{#3}%
  \fi%
}

\acrodef{GPE}{Gross-Pitaevskii equation}
\acrodef{BH}{Bose-Hubbard}
\acrodef{MF}{mean-field}
\acrodef{BBR}{Bogoliubov back-reaction}
\acrodef{BGL}{balanced gain and loss}
\acrodef{BEC}{Bose-Einstein condensate}
\acrodef[longplural={single-particle density matrices}]{SPDM}{single-particle density matrix}
\acrodef[longplural={two-particle density matrices}]{TPDM}{two-particle density matrix}

\begin{document}

% Use the \preprint command to place your local institutional report
% number in the upper righthand corner of the title page in preprint mode.
% Multiple \preprint commands are allowed.
% Use the 'preprintnumbers' class option to override journal defaults
% to display numbers if necessary
\preprint{}

\begin{abstract} A quantum system exhibiting $\mathcal{PT}$ symmetry is a
Bose-Einstein condensate in a double-well potential with balanced particle gain
and loss, which is described in the mean-field limit by a Gross-Pitaevskii
equation with a complex potential. A possible experimental realization of such a
system by embedding it into a Hermitian time-dependent four-mode optical lattice
was proposed by Kreibich \textit{et al.}\
[\href{https://doi.org/10.1103/PhysRevA.87.051601}{Phys.\ Rev.\ A \textbf{87},
051601(R) (2013)}], where additional potential wells act as reservoirs and
particle exchange happens via tunneling. Since particle influx and outflux have
to be controlled explicitly, a set of conditions on the potential parameters had
to be derived. In contrast to previous work, our focus lies on a full many-body
description beyond the mean-field approximation using a Bose-Hubbard model with
time-dependent potentials. This gives rise to novel quantum effects, such that
the differences between mean-field and many-body dynamics are of special
interest. We further present stationary \emph{analytical} solutions for the
embedded wells in the mean-field limit, different approaches for the embedding
into a many-body system, and a very efficient method for the evaluation of
hopping terms to calculate exact Bose-Hubbard dynamics.  \end{abstract}

\title{Realization of balanced gain and loss in a time-dependent four-mode Bose-Hubbard model}
\author{Daniel Dizdarevic}
\email{daniel.dizdarevic@itp1.uni-stuttgart.de}
\author{J\"org Main}
\author{Kirill Alpin}
\author{Johannes Reiff}
\author{Dennis Dast}
\author{Holger Cartarius}
\author{G\"unter Wunner}
\affiliation{Institut f\"ur Theoretische Physik 1, Universit\"at Stuttgart, 70550 Stuttgart, Germany}
\date{\today}
\maketitle

\section{Introduction}
\label{sec:introduction}

In recent years a growing interest in non-Hermitian and especially
$\PT$-symmetric quantum mechanics has arisen, since such theories allow for an
effective description of open quantum systems \cite{Moiseyev2011}. $\PT$
symmetry \cite{Bender1998}, meaning that the Hamiltonian is invariant under
combined parity and time reflection, has already been observed in optical
systems \cite{Guo2009,Rueter2010,Peng2014,Weimann2017}, microwave cavities
\cite{Bittner2012} and electronics \cite{Schindler2011}.

A well investigated and experimentally accessible genuine quantum system
exhibiting $\PT$ symmetry was proposed by Klaiman \textit{et al.}
\cite{Klaiman2008} and is given by a \ac{BEC} in a double-well potential with
\ac{BGL}
\cite{Dast2013a,Dast2013b,Haag2014,Dizdarevic2015,Graefe2012,Shin2005,Gati2006}.
In theory, gain and loss can be introduced via complex potentials
\cite{Kagan1998} or Lindblad superoperators \cite{Lindblad1976}. Although such
descriptions render the Hamiltonian non-Hermitian, in the $\PT$-symmetric case
stationary states with real eigenvalues exist.

Experimental techniques for the realization of gain and loss are well developed
these days. Localized gain can be realized by feeding atoms from a second
condensate \cite{Robins2008,Doering2009}, while localized loss can be achieved
with a focused electron beam \cite{Gericke2008,Wuertz2009,Barontini2013}.
Although the experimental tools for a direct realization of $\PT$-symmetric
\acp{BEC} are in principle available, no such experiment has been accomplished
so far. Other realizations of $\PT$-symmetric systems in ultracold atoms were
proposed by means of bound wave functions \cite{Single2014,Gutoehrlein2015}, and
recently experiments were performed using only localized loss, which, however,
effectively show \ac{BGL} \cite{Labouvie2016}. There are also theoretical
investigations of the double-well system beyond a Gross-Pitaevskii dynamics,
where the open quantum system is described by a Lindblad-type master equation
\cite{Dast2014,Dast2016b}.

Here, however, we will focus on yet another approach that is an embedding of a
two-mode \ac{BEC} in a larger Hermitian system \cite{Kreibich2013,Kreibich2014}
allowing for a causal description of particle gain and loss via tunneling. To
this end, parameters of the optical lattice have to be varied in a
time-dependent manner based on the current state of the system. While this
method is already theoretically well investigated within the \ac{MF}
approximation even for large particle numbers \cite{Kreibich2016}, which is in
principle sufficient for the description of a potential experiment, we are
particularly interested in a many-body description of such systems. By
consideration of two-mode models novel generic quantum effects in open many-body
systems were found, e.g.\ purity oscillations \cite{Dast2016a,Dast2016b}, which
are experimentally manifested in the average contrast \cite{Dast2017}. To
experimentally investigate such effects \ac{BGL} has to be established over
sufficiently long timescales. This makes a many-body realization of the two-mode
model in the low particle number limit of particular interest, which will be the
main topic of this paper.

This paper is organized as follows. We will start by giving a short overview of
the \ac{BH} model and a discussion about different approximations in
Sec.~\ref{sec:MB_models}. Afterwards, in Sec.~\ref{sec:realization}, we will
introduce the $\PT$-symmetric two-mode model and its embedding into a four-mode
optical lattice in the \ac{MF}. We will transfer these results to many-body
systems in Sec.~\ref{sec:results} by also employing time-dependent potentials in
the \ac{BH} model. It will be discussed in detail whether $\PT$ symmetry or
\ac{BGL} as known from the \ac{MF} can still be observed in a double-well system
embedded into an optical lattice by using a full many-body description. Finally,
we give some conclusions and an outlook in Sec.~\ref{sec:conclusions}.

\section{Models and methods for many-body systems}
\label{sec:MB_models}

In this section we will recapitulate the basic concepts and methods necessary
for a many-body description of optical lattice systems. The equations discussed
here are valid for any $M$-mode optical lattice and lay the foundation for the
following sections, where we will consider the special case of four-mode optical
lattice systems.

\subsection{Bose-Hubbard model}
\label{subsec:MB_models/BHM}

The \ac{BH} model \cite{Gersch1963} is a model for the description of the
many-body dynamics of trapped bosons in multi-well potentials, such as ultracold
atoms in optical lattices \cite{Jaksch1998}. The \ac{BH} Hamiltonian for an
$M$-well system reads
\begin{align}
  \Hamiltonian_\BH(t) &= -\sum_{\mathclap{\angles{m,m'}}} J_{mm'}(t) \ad_m\a_{m'}
    + \frac{1}{2} \sum_m U_m(t) \n_m \parens*{\n_m - 1} \nonumber \\
  &\quad+ \sum_m \mu_m(t) \n_m \comma \label{eq:BHM}
\end{align}
where $\angles{\cdot,\cdot}$ denotes the summation of all nearest neighbors for
$m, m' \in \brackets{1, M}$. The operators $\ad_m$ and $\a_m$ are the bosonic
particle creation and annihilation operators for well $m$ and $\n_m = \ad_m\a_m$
denotes the corresponding particle number operator.

The first term is the kinetic part of the \ac{BH} Hamiltonian, which describes
particle exchange between nearest neighbors with rates $J_{mm'}$, whereas the
second and third terms are the potential energies due to contact interactions
$U_m$ and chemical potentials $\mu_m$ in each well, respectively; the last term
in Eq.~\eqref{eq:BHM} is therefore also called \emph{on-site energy}.

It should be noted at this point, that the parameters in our model are
explicitly time-dependent in contrast to most other applications. This is due to
the fact, that we will later use the potential parameters $J_{mm'}$ and $\mu_m$
to control the dynamics in the system.  Thus, the interaction energy $U_m$ is
time-dependent as well, since it involves an overlap integral of the wave
function at one site \cite{Jaksch1998}. However, for the sake of simplicity we
will only consider the case where $U_m(t) = U$ is constant and equal at each
site. Considering only s-wave scattering, where the parameter $U$ depends on the
scattering length $a$ as described in Ref.\ \cite{Jaksch1998}, this assumption
corresponds to a dynamical $a(t)$, which is realizable by Feshbach resonances
\cite{Feshbach1958,Chin2010}. Nevertheless, the method presented in the
following sections is still applicable in the general case of Eq.\
\eqref{eq:BHM} by appropriate modifications, though, no change in the
qualitative behavior is expected.

A natural basis for the $M$-mode \ac{BH} model is formed by the Fock states
$\ket{n_1, \ldots, n_M}$ with $\sum_{k = 1}^M n_k = N$. Although our \ac{BH}
model \eqref{eq:BHM} explicitly contains time-dependent potentials we can still
resort to standard methods to calculate its exact time evolution and dynamics
(see e.g.\ \cite{Zhang2010}) while using a slightly advanced method for
evaluating the hopping terms. After a pairwise application of creation and
annihilation operators, we calculate the position of the resulting vector in a
lexicographically ordered basis directly as shown in the Appendix. Since most
matrix elements contain time-dependent parameters and therefore depend on the
current state of the system, matrix multiplications have to be evaluated on the
fly.

\subsection{Approximations to the Bose-Hubbard model}
\label{subsec:MB_models/BBR}

The dimension of the Fock basis scales roughly as $N^{M-1}$ for $N \gg 1$ (cf.\
Eq.~\eqref{eq:dimension}) and grows rapidly with an increasing number of
particles, thus restricting the numerical computations for exact many-body
dynamics to particle numbers $N \lesssim 200$. To enter the larger particle
number regime, in the following we will briefly discuss different approximations
to the \ac{BH} model.

The simplest approximation possible is the well-known \ac{MF} approximation (see
e.g.\ Ref.\ \cite{Pisarski2011}), where the dynamics is given by the average
behavior of all particles in the ground state.  With the notation $\psi_m =
\bra{\psi_0} \a_m \ket{\psi_0}$ and for $N \gg 1$ the Heisenberg equations of
motion for annihilation and creation operators yield \begin{align} \imag
\pdiff{}{t} \psi_m &= -J_{m,m-1} \psi_{m-1} - J_{m,m+1} \psi_{m+1} \nonumber\\
&\quad + U_m n_m \psi_m + \mu_m \psi_m \comma \label{eq:discrete_GPE}
\end{align} which is the discrete \ac{GPE} \cite{Gross1961,Pitaevskii1961} also
obtainable from the continuous \ac{GPE} via a frozen Gaussian ansatz
\cite{Heller1981,Huber1986,Graefe2008,Kreibich2014}.  This \ac{MF} approximation
describes the dynamics of a \ac{BEC} in the limit $N \rightarrow \infty$ and $T
\rightarrow 0$, where the occupation of the ground state is dominant. However,
there is a huge downside; every particle is assumed to be in the same
single-particle state and thus all many-particle information about the
interacting system is lost completely.

An approximation beyond the \ac{MF} for the treatment of large numbers of
particles is the so-called \ac{BBR} method
\cite{Anglin2001a,Anglin2001b,Tikhonenkov2007}. By using the density operator
$\operator{\rho}$, the full many-body dynamics described by the \ac{BH} model
\eqref{eq:BHM} can be rewritten in terms of the expectation values of the
entries of the \ac{SPDM} $\spdm_{ij} = \expval{\ad_i \a_j}$, the \ac{TPDM}
$\tpdm_{ijkl} = \expval{\ad_i \a_j \ad_k \a_l}$, or in general the $n$-particle
density matrices
\begin{equation}
  \xpdm{n}_{i_1 \ldots i_n} = \expval{\ad_{i_1} \a_{i_1} \ldots \ad_{i_n} \a_{i_n}} \period
  \label{eq:density_matrix}
\end{equation}
The first two orders of this BBGKY-type hierarchy \cite{Anglin2001b} read
\begin{subequations}
\begin{align}
  \imag \pdiff{}{t} \spdm_{ij} &= \Zeta{1}_{ij}
    - \parens*{\mu_i - \mu_j} \spdm_{ij} \comma \\
  \imag \pdiff{}{t} \tpdm_{ijkl} &= \Zeta{2}_{ijkl}
    - \parens*{\mu_i - \mu_j
    + \mu_k - \mu_l} \tpdm_{ijkl} \comma
\end{align}
where, for later purposes, we define
\begin{align}
  \Zeta{1}_{ij} &= J_{i-1,i} \spdm_{i-1,j} + J_{i+1,i} \spdm_{i+1,j} \nonumber\\
  &\quad - J_{j,j-1} \spdm_{i,j-1} - J_{j,j+1} \spdm_{i,j+1} \nonumber\\
  &\quad - U \parens*{\tpdm_{iiij} - \tpdm_{ijjj}} \comma \label{eq:SPDM}\\
  \Zeta{2}_{ijkl} &= J_{i-1,i} \tpdm_{i-1,jkl} + J_{i+1,i} \tpdm_{i+1,jkl} \nonumber\\
  &\quad - J_{j,j-1} \tpdm_{i,j-1,kl} - J_{j,j+1} \tpdm_{i,j+1,kl} \nonumber\\
  &\quad + J_{k-1,k} \tpdm_{ij,k-1,l} + J_{k+1,k} \tpdm_{ij,k+1,l} \nonumber\\
  &\quad - J_{l,l-1} \tpdm_{ijk,l-1} - J_{l,l+1} \tpdm_{ijk,l+1} \nonumber\\
  &\quad - U \parens*{\xpdm{3}_{iiijkl} - \xpdm{3}_{ijjjkl} + \xpdm{3}_{ijkkkl}
    - \xpdm{3}_{ijklll}} \period \label{eq:TPDM}
\end{align}
\label{seq:BBGKY}%
\end{subequations}
It should be noted that the $n$th order in this hierarchy is coupled to the
$(n+1)$st order by the contact interaction strength, where $n < N$, with the
total number of particles $N$ in the system.

Now the \ac{BBR} approximation \cite{Anglin2001a,Anglin2001b,Tikhonenkov2007}
comes into play cutting off the hierarchy \eqref{seq:BBGKY} after the
second-order such that a closed system of coupled differential equations is
formed. Following \cite{Tikhonenkov2007}, we start with
\begin{equation}
  \ospdm_{ij} \approx \spdm_{ij} + \delta\ospdm_{ij} \period
  \label{eq:ansatz_SPDM}
\end{equation}
To uncouple the Eqs.~\eqref{seq:BBGKY} from the remaining dynamics, one has to
expand $\xpdm{3}_{ijklmn}$ using Eq.~\eqref{eq:ansatz_SPDM}, which yields
\begin{align}
  \xpdm{3}_{ijklmn} &\approx \spdm_{ij}\tpdm_{klmn} + \spdm_{mn}\tpdm_{ijkl}
    + \spdm_{kl}\tpdm_{ijmn} \nonumber\\
  &\quad - 2\spdm_{ij}\spdm_{kl}\spdm_{mn} \period \label{eq:BBR_2}
\end{align}
Apart from the single-particle dynamics, this \ac{BBR} method also contains
two-particle dynamics, which is the most significant many-body contribution
beyond the \ac{MF} dynamics. In principle, even better \ac{BBR} approximations
are possible by cutting off the hierarchy at higher orders; however, the
computational cost increases exponentially.

The \ac{BBR} approximation and the BBGKY hierarchy in general feature a specific
symmetry, that is
\begin{equation}
  \conjg{\parens*{\xpdm{n}_{k \cdots l}}} = \xpdm{n}_{l \cdots k} \comma
  \label{eq:BBR_symm}
\end{equation}
i.e., a complex conjugation will reverse the order of the indices. This comes in
handy to reduce computational effort to some extent.

\section{Realization of the two-mode model in the mean-field}
\label{sec:realization}

In the following section we will introduce the $\PT$-symmetric two-mode model in
the \ac{MF} approximation. Then, the basic concept of embedding this open
quantum system into a larger Hermitian system is discussed for a four-mode
system. We derive fundamental conditions for the embedding and analytical
solutions for stationary states.

\subsection{$\PT$-symmetric two-mode model}
\label{sec:realization/TMM}

The dynamics of a \ac{BEC} at very low temperatures and in the limit of large
particle numbers is well described by the \ac{GPE}. If the wells of the trapping
potential are sufficiently deep, the \ac{GPE} can be reduced to a matrix model
(cf.\ Eq.\ \eqref{eq:discrete_GPE}) \cite{Graefe2012,Kreibich2013,Kreibich2014}.
In the case of a $\PT$-symmetric double-well potential \cite{Dast2013a}, the
dimensionless discrete \ac{GPE} reads \cite{Graefe2012}
\begin{equation}
  \imag \pdiff{}{t}
    \begin{pmatrix}
      \psi_1 \\
      \psi_2
    \end{pmatrix}
  = \begin{pmatrix}
      g \absquare{\psi_1} + \imag \gamma & -J_{12} \\
      -J_{12} & g \absquare{\psi_2} - \imag \gamma
    \end{pmatrix}
    \begin{pmatrix}
      \psi_1 \\
      \psi_2
    \end{pmatrix} \comma
    \label{eq:TMM}
\end{equation}
where
\begin{equation}
  \psi_k = \sqrt{n_k}\exp\parens{\imag\phi_k}
  \label{eq:MF_wave_function}
\end{equation}
are complex numbers describing the \ac{MF} wave function at site $k$. The
macroscopic interaction strength
\begin{equation}
  g = (N-1)U
  \label{eq:g_U}
\end{equation}
describes contact interactions for $N$ particles with interaction energy $U$ as
in Eq.~\eqref{eq:BHM} and vanishes for a single particle, i.e., $N = 1$. In the
\ac{MF}, i.e., for a large number of particles, $(N-1) \sim N$ can be used; we
are, however, interested in the low particle number limit. Particle exchange
between neighboring sites is given by the off-diagonal elements of the
Hamiltonian with tunneling constants $J_{12} \ge 0$, while particle exchange
with the environment is described by an antisymmetric imaginary potential with
coupling strength $\gamma$. This non-Hermitian two-mode model on the one hand
shows $\PT$-symmetric stationary states with completely real eigenvalues and
non-zero current, and on the other hand $\PT$-broken states with complex
eigenvalues \cite{Graefe2012} as well as rich bifurcation schemes
\cite{Dast2013a,Dast2013b,Haag2014,Dizdarevic2015}. For sufficiently small
values of $\gamma$ and $g$, the eigenvalue spectrum is real and there exist
$\PT$-symmetric stationary solutions to Eq.~\eqref{eq:TMM}.

\subsection{Embedding into a four-mode system}
\label{subsec:realization/embedding}

\begin{figure}
  \includegraphics[width=\columnwidth]{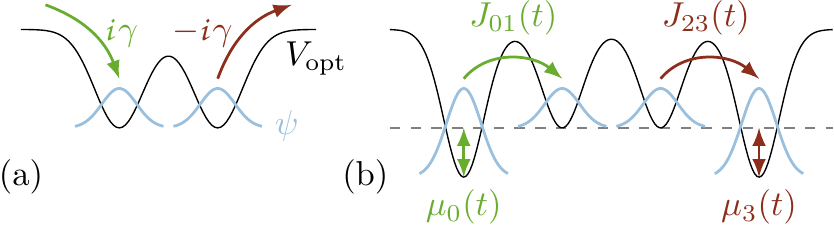}
  \caption{Open non-Hermitian two-mode system (a) and closed Hermitian four-mode
  system (b). The currents between the outer reservoir wells and the inner wells
  account for the in- and out-coupling of particles in the open quantum system.
  To reproduce the properties of an open system the additional parameters
  $J_{01}(t)$, $J_{23}(t)$, $\mu_0(t)$ and $\mu_3(t)$ have to be varied
  time-dependently.}
  \label{fig:TMM_FMM}
\end{figure}

Although the $\PT$-symmetric two-mode model \eqref{eq:TMM} is quite simple, so
far there was no observation of $\PT$ symmetry in a real quantum system because
such an open quantum system is hard to realize experimentally. For this reason,
we rely on \cite{Kreibich2013} and, as shown in Fig.~\ref{fig:TMM_FMM}, embed
the two-mode model into a four-mode optical potential, which can be painted in a
time-averaged manner using a rapidly moving laser beam \cite{Henderson2009}.
Although there is a method available for creating an actual loss as described by
an imaginary potential $-\imag\gamma$ by using an electron beam
\cite{Barontini2013,Santra2015,Labouvie2016}, we want to focus completely on the
realization of an open quantum system via the approach of embedding it into a
larger structure, where the reservoir is explicitly modeled and controlled.

By embedding a two-mode system into a larger Hermitian system, we allow for
particle and energy exchange between the two-mode sub-system, in the following
simply referred to as the \emph{embedded system}, and the additional
\emph{reservoir wells}. While one additional well at each side of the system is
sufficient for the embedding, such a reservoir will in general not fulfill the
Markov property. Markovian behavior, i.e., memorylessness of the environment is
a fundamental assumption underlying the $\PT$-symmetric two-mode model and its
many-body counterpart given by a Lindblad-type master equation. For large
population numbers, however, the finite size of the reservoir becomes less
significant. Systems with a large number of reservoir wells were investigated in
\cite{Kreibich2016} for the \ac{MF} case.

While also the four-mode system has already been studied intensively
\cite{Kreibich2013,Kreibich2014}, we will not only recap these previous results,
but rather rework the \ac{MF} problem in order to lay the foundation for
Sec.~\ref{sec:results}. Although those previous results are in principle
applicable to many-body systems, their complexity provides difficulties.
Therefore, it is required to discuss the \ac{MF} theory, although our intention
is to go beyond \ac{MF} in the present paper. Moreover, by directly using the
equations of motion for the \acp{SPDM} $\sigma_{ij} = \conjg{\psi}_i \psi_j$,
which will be seen later is a natural approach to also employ the results in
many-body calculations, a degree of freedom $d(t)$ remains, i.e., an arbitrary
time-dependent function, which does not change the observables at all. We,
however, eliminate this degree of freedom by using the equations of motion for
the macroscopic wave function $\psi$ instead, which will also yield much simpler
results in terms of complexity as will be shown in this section. For those who
are familiar with Ref.\ \cite{Kreibich2014}, though, we want to mention that our
results are in fact fully equivalent to their results for the unique functional
relation
\begin{equation}
  d(t) = -\frac{J_{12}}{2 \sqrt{n_0(t) n_3(t)}} \period
  \label{eq:freedom}
\end{equation}

To emulate the in- and out-coupling of particles in analogy with the effects of
a complex potential in the embedded system, we have to control the tunneling
processes by altering the additional parameters of the optical potential in the
reservoir wells, namely $J_{01}(t)$, $J_{23}(t)$, $\mu_0(t)$ and $\mu_3(t)$, in
a time-dependent manner. To this end, the dynamics of the $\PT$-symmetric
two-mode model \eqref{eq:TMM} is compared to the embedded dynamics in a
time-dependent four-mode model with the Hamiltonian
\begin{equation}
  \Hamiltonian_4 =
  \begin{psmallmatrix}
    g \absquare{\psi_0} + \mu_0(t) & -J_{01}(t) & 0 & 0 \\
    -J_{01}(t) & g \absquare{\psi_1} & -J_{12} & 0 \\
    0 & -J_{12} & g \absquare{\psi_2} & -J_{23}(t)  \\
    0 & 0 & -J_{23}(t) & g \absquare{\psi_3} + \mu_3(t)
  \end{psmallmatrix} \comma
  \label{eq:FMM}
\end{equation}
which immediately yields the complex-valued conditions
\begin{subequations}
\begin{align}
  J_{01} \psi_0 &= -\imag\gamma \psi_1 \comma \label{eq:condition_MF_1} \\
  J_{23} \psi_3 &= \imag\gamma \psi_2 \period \label{eq:condition_MF_2}
\end{align}
\label{seq:condititions_MF}%
\end{subequations}
A remarkable feature of these conditions is the necessity of a defined
antisymmetric phase relation between system and reservoir wells,
\begin{subequations}
\begin{align}
  \phi_0 - \phi_1 &= -\frac{\pi}{2} \comma \label{eq:dphi01} \\
  \phi_3 - \phi_2 &= \frac{\pi}{2} \comma \label{eq:dphi23}
\end{align}
\label{seq:dphi}%
\end{subequations}
which is a consequence of the definition \eqref{eq:MF_wave_function}. If these
phase relations hold, in- and out-coupling of particles can indeed be realized
by particle currents and the time-dependent tunneling rates are given by
\begin{subequations}
\begin{align}
  J_{01}(t) &= \frac{2 \gamma n_1(t)}{\jt_{01}(t)} \comma \label{eq:J01_MF} \\
  J_{23}(t) &= \frac{2 \gamma n_2(t)}{\jt_{23}(t)} \comma \label{eq:J23_MF}
\end{align}
\label{seq:J_MF}%
\end{subequations}
where we introduced the reduced current density
\begin{equation}
  \jt_{ij} = 2\ipart\parens{\conjg{\psi}_i\psi_j}
  \label{eq:reduced_current}
\end{equation}
such that $j_{ij} = J_{ij}\jt_{ij}$ describes the current between wells $i$ and
$j$. Analogously we also introduce
\begin{equation}
  c_{ij} = 2\rpart\parens{\conjg{\psi}_i\psi_j} \propto \cos\parens{\phi_i - \phi_j}
  \label{eq:correlations}
\end{equation}
such that with Eqs.~\eqref{seq:dphi}
\begin{equation}
  c_{01} = c_{23} = 0 \period
  \label{eq:c_eq_0}
\end{equation}

As the previous considerations show, the embedding critically depends on the
conservation of specific phase relations. To ensure the validity of
Eqs.~\eqref{seq:dphi}, it is sufficient to demand that they hold initially and
$\dot{c}_{01} = \dot{c}_{23} = 0$ for all times $t$, which yields
\begin{subequations}
\begin{align}
  \mu_0(t) &= -J_{12} \frac{\jt_{02}(t)}{\jt_{01}(t)}
    - g\parens*{n_0(t) - n_1(t)} \comma \label{eq:mu0_MF} \\
  \mu_3(t) &= -J_{12} \frac{\jt_{13}(t)}{\jt_{23}(t)}
    - g\parens*{n_3(t) - n_2(t)} \period \label{eq:mu3_MF}
\end{align}
\label{seq:mu_MF}%
\end{subequations}
With the time-dependent parameters \eqref{seq:J_MF} and \eqref{seq:mu_MF}, the
dynamics of the embedded wells resembles the dynamics of the open quantum system
\eqref{eq:TMM} for all initial states satisfying the phase relations
\eqref{seq:dphi}, which is in agreement with the previous results from
\cite{Kreibich2013,Kreibich2014} using Eq.~\eqref{eq:freedom}.

In the following, we will consider quasi-stationary states of the four-mode
model, i.e., steady states of the embedded system corresponding to
$\PT$-symmetric states in the two-mode model \cite{Graefe2012}. Such states
arise from the initial conditions
\begin{subequations}
\begin{align}
  n_1(0) = n_2(0) &= n \comma \label{eq:init_n} \\
  j_{01}(0) = j_{12}(0) = j_{23}(0) &= 2 \gamma n \period \label{eq:init_j}
\end{align}
\label{seq:embedded_init}%
\end{subequations}
Without loss of generality, the initial conditions for the \ac{MF} wave function
read
\begin{subequations}
\begin{align}
  \psi_0(0) &= \sqrt{n_0(0)} \exp\parens{\imag\parens*{\phi - \pi/2}} \comma
    \label{eq:init_psi0} \\
  \psi_1(0) &= \sqrt{n} \exp\parens{\imag\phi} \comma \label{eq:init_psi1} \\
  \psi_2(0) &= \sqrt{n} \exp\parens{-\imag\phi} \comma \label{eq:init_psi2} \\
  \psi_3(0) &= \sqrt{n_3(0)} \exp\parens{-\imag\parens*{\phi - \pi/2}}
    \label{eq:init_psi3}
\end{align}
with
\begin{equation}
  \phi = -\frac{1}{2} \arcsin\parens*{\frac{\gamma}{J_{12}}} \period
  \label{eq:init_phi_MF}
\end{equation}
\label{seq:init_MF}%
\end{subequations}
If we demand constant gain and loss, i.e., $\gamma$ is constant, the conditions
\eqref{seq:embedded_init} are preserved for all times. Therefore, the occupation
numbers in the reservoir will change linearly in time,
\begin{subequations}
\begin{align}
  n_0(t) &= n_0(0) - 2 \gamma n t \comma \label{eq:stat_n0_MF} \\
  n_3(t) &= n_3(0) + 2 \gamma n t \period \label{eq:stat_n3_MF}
\end{align}
\label{seq:stat_n_MF}%
\end{subequations}
The Eqs.~\eqref{seq:dphi}, \eqref{seq:embedded_init} and \eqref{seq:stat_n_MF}
can be used to obtain analytical expressions for the time-dependent lattice
parameters,
\begin{subequations}
\begin{align}
  J_{01}(t) &= \gamma \sqrt{\frac{n}{n_0(0) - 2 \gamma n t}} \comma \label{eq:J01_analyt} \\
  J_{23}(t) &= \gamma \sqrt{\frac{n}{n_3(0) + 2 \gamma n t}} \comma \label{eq:J23_analyt} \\
  \mu_0(t) &= \mu - g \parens{n_0(0) - 2 \gamma n t} \comma \label{eq:mu0_analyt} \\ 
  \mu_3(t) &= \mu - g \parens{n_3(0) + 2 \gamma n t} \comma \label{eq:mu3_analyt}
\end{align}
where the chemical potential $\mu$ is given by
\begin{equation}
  \mu = g n - \sqrt{J_{12}^2 - \gamma^2} \period
  \label{eq:chem_pot}
\end{equation}
\label{seq:parameters_MF}
\end{subequations}
Note that these analytical results go beyond \cite{Kreibich2013,Kreibich2014}.

A remarkable feature of Eqs.~\eqref{seq:parameters_MF} is that initial
populations in the reservoir wells can be chosen arbitrarily. However, the
choice of $n_0(0)$ determines the time scale on which quasi-stationary states
can be realized, i.e., the system breaks down when the left reservoir is empty
at time
\begin{equation}
  \tau = \frac{n_0(0)}{2 \gamma n} \period
  \label{eq:time_scale}
\end{equation}

To summarize, by means of time-dependent parameters, stationary states and
therefore $\PT$ symmetry can be realized in the embedded system for $0 \le t \le
\tau$ within the \ac{MF} approximation, i.e., the inner wells of the Hermitian
four-mode system \eqref{eq:FMM} behave exactly as the non-Hermitian two-mode
system \eqref{eq:TMM}.

\section{Results and discussion}
\label{sec:results}

It was shown in Sec.~\ref{sec:realization}, that the $\PT$-symmetric two-mode
model can be realized in the \ac{MF} using a time-dependent four-mode optical
lattice. Naturally the question arises, if this still holds in the limit of low
particle numbers, i.e., when a full many-body description is required.
Therefore, two different approaches are presented, where the previous results
are applied to the many-body four-mode system in order to investigate if
realizations of the two-mode model and $\PT$ symmetry are also feasible beyond
the \ac{MF} approximation.

\subsection{Pure many-body states}
\label{subsec:results/pure_states}

First, we have to create the same initial condition for a many-body system, that
is we construct a pure many-body state from a \ac{MF} state
\eqref{eq:MF_wave_function}. To do so, an ansatz of identical single-particle
wave functions is used, which can be written in terms of Fock states as
\begin{subequations}
\begin{equation}
  \ket{\psi, N} = \smashoperator{\sum_{n_1, \ldots, n_M}}
    u_{n_1, \ldots, n_M} \ket{n_1, \ldots, n_M}
  \label{eq:pure_wavefunction}
\end{equation}
with the sum running over all possible combinations with total particle number
$N$. The coefficients are then given by
\begin{equation}
  u_{n_1, \ldots, n_M} = \sqrt{N!} \prod_{m = 1}^M
    \frac{\psi_m^{n_m}}{\sqrt{n_m!}} \comma \label{eq:coefficients}
\end{equation}
\label{seq:pure_state_BH}%
\end{subequations}
where each component of the macroscopic wave function $\psi_m$ is raised to the
power of the corresponding population and normalized appropriately.

To obtain the representation of a pure many-body state also for the \ac{BBR}
approximation, we consider the action of the annihilation operator onto the
state \eqref{eq:pure_wavefunction},
\begin{equation}
  \a_k \ket{\psi, N} = \sqrt{N} \psi_k \ket{\psi, N-1} \comma
  \label{eq:annihil_pure_state}
\end{equation}
which again yields a pure many-body state for $N-1$ particles. This leads to
\begin{subequations}
\begin{align}
  \spdm_{ij} &= N \conjg{\psi}_i \psi_j \comma \label{eq:pure_SPDM} \\
  \tpdm_{ijkl} &= N(N-1) \conjg{\psi}_i \psi_j \conjg{\psi}_k \psi_l
    + N \conjg{\psi}_i \delta_{jk} \psi_l \period \label{eq:pure_TPDM}
\end{align}
\label{seq:pure_state_BBR}%
\end{subequations}

With Eqs.~\eqref{seq:pure_state_BH} and \eqref{seq:pure_state_BBR} we can now
prepare the system in a pure state corresponding to \ac{MF} states as in
Eq.~\eqref{eq:MF_wave_function}. In the following sections we will always assume
that the four-mode system is prepared in an initial state corresponding to
Eqs.~\eqref{seq:init_MF}. Such an initial state corresponds indeed to the ground
state of the respective embedded two-mode system in the \ac{MF}, but is in
general not the ground state of the many-body system itself.

\subsection{Mean-field-like approach}
\label{subsec:results/conditions_beyond_MF}

The results of Sec.~\ref{sec:realization} are in agreement with Refs.\
\cite{Kreibich2013,Kreibich2014} together with Eq.~\eqref{eq:freedom}. We now
want to go beyond the \ac{MF} approximation using the methods discussed in
Sec.~\ref{sec:MB_models} for $M = 4$ in order to examine which many-body effects
occur and whether $\PT$-symmetric stationary states can still be established or
not. Since the \ac{BH} model contains more degrees of freedom than the \ac{GPE},
whereas there are at the same time no additional free control parameters, that
problem is highly nontrivial.

There is another important difference between \ac{MF} and many-body dynamics we
have to pay special attention to. Calculations for the two-mode model in the
\ac{MF} approximation are typically performed using $n_1 = n_2 = 1/2$. The
reason is that the \ac{GPE} only describes a single-particle dynamics and the
total particle number $N_2 = n_1 + n_2$ is just a parameter, which can therefore
be normalized to $N_2 = 1$. However, for the \ac{BH} model and the \ac{BBR}
approximation, this is not the case anymore. To allow for a comparison with the
open two-mode system \eqref{eq:TMM}, which is described in terms of the
(normalized) macroscopic interaction strength $g$, the macroscopic interaction
strength in the four-mode model $g_4$ has to be modified according to
\begin{equation}
  g_4 = \frac{g}{N_2} \period
  \label{eq:modified_g}
\end{equation}
The contact interaction potential $U$ for many-body calculations is then
obtained from $g_4$ by using Eq.~\eqref{eq:g_U}, that is $U = g_4/(N-1)$.

Let us start in a pure many-body state \eqref{seq:pure_state_BH} corresponding
to Eqs.~\eqref{seq:init_MF}, i.e., in particular the relations \eqref{seq:dphi}
hold. We then introduce time-dependent parameters also in the \ac{BH} model as
in Sec.~\ref{sec:realization}. As a first attempt we try to impose the same
conditions as in the \ac{MF}. Thus, we can directly check the effects of
many-particle corrections on the dynamics and whether $\PT$ symmetry can be
implemented using exactly the same conditions as in the \ac{MF}.

First, we demand that the currents are given by the coupling strength $\gamma$
which gives formally the same results for the tunneling rates as in
Sec.~\ref{subsec:realization/embedding}, which implies that
Eqs.~\eqref{seq:J_MF} are also usable for the \ac{BH} model with $\jtx{1}_{ij} =
2\ipart{\spdm_{ij}}$.

Second, we evaluate $\dot{c}^{(1)}_{01} = \dot{c}^{(1)}_{23} = 0$ for the
many-body dynamics where $\cx{1}_{ij} = 2 \rpart{\spdm_{ij}}$. This yields
\begin{subequations}
\begin{align}
  \mu_0(t) &= \frac{\Ups{1}_{01}(t)}{\jtx{1}_{01}(t)} \nonumber\\
  &= -J_{12} \frac{\jtx{1}_{02}(t)}{\jtx{1}_{01}(t)} - U \frac{\jtx{2}_{0001}(t)
    - \jtx{2}_{0111}(t)}{\jtx{1}_{01}(t)} \comma \label{eq:mu0_MB}\\
  \mu_3(t) &= \frac{\Ups{1}_{23}(t)}{\jtx{1}_{23}(t)} \nonumber\\
  &= -J_{12} \frac{\jtx{1}_{13}(t)}{\jtx{1}_{23}(t)} + U \frac{\jtx{2}_{2223}(t)
    - \jtx{2}_{2333}(t)}{\jtx{1}_{23}(t)} \comma \label{eq:mu3_MB}
\end{align}
where
\begin{equation}
  \Ups{1}_{ij} = 2\ipart{\Zeta{1}_{ij}} \period
  \label{eq:Ups}
\end{equation}
\label{seq:mu_MB}%
\end{subequations}
The first-order terms are now marked explicitly, since second-order terms
corresponding to \acp{TPDM} are present, that is $\jtx{2}_{ijkl} =
2\ipart{\tpdm_{ijkl}}$. In contrast to the \ac{MF} parameters \eqref{seq:mu_MF},
the above equations contain corrections of second-order due to a coupling of the
equations of motion via interactions. At this point it should also be noted that
the choice of parameters \eqref{seq:mu_MB} can no longer fulfill the relations
\eqref{seq:dphi} except for perfectly pure states, that is only in the \ac{MF}.
Therefore, this approach should be viewed as an extension of the \ac{MF} method
in the vicinity of pure states.

\begin{figure*}
  \includegraphics[width=0.9\textwidth]{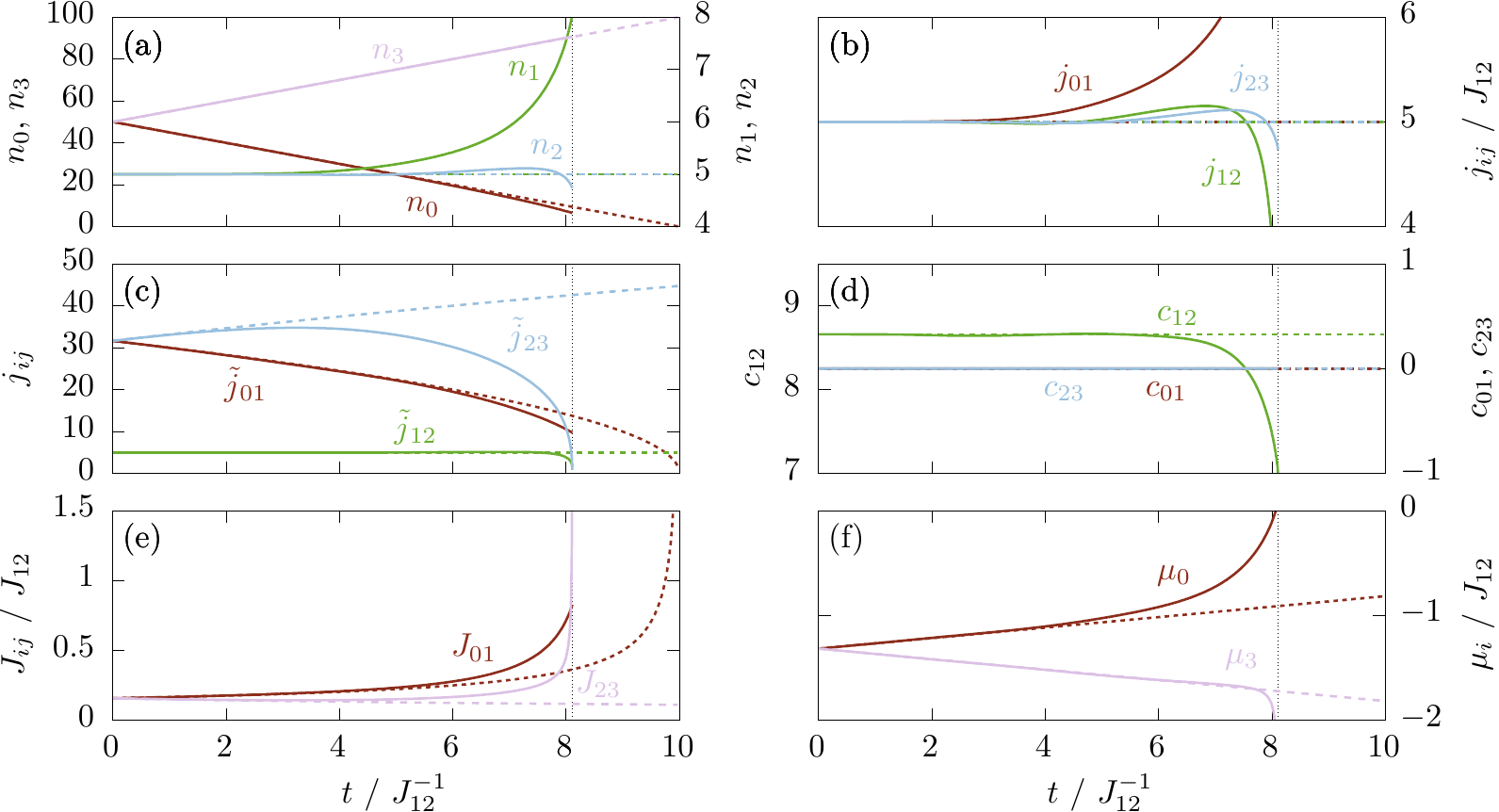}%
  \caption{Observables and parameters for a weakly interacting \ac{BEC} with $g
  = 0.1$ described by the \ac{BH} model. The coupling strength is fixed to
  $\gamma = 0.5$ and the occupation ratio between the embedded system and the
  reservoir is $10$ for a total number of $N = 110$ particles. In comparison
  with the \ac{MF} limit (dashed lines), the many-body system (solid lines)
  shows no quasi-stationary behavior, which is clearly visible by looking at the
  occupation numbers $n_1$ and $n_2$ (a), and the reduced current $\jt_{12}$ (b)
  and $c_{12}$ (d). The many-body calculations also show an earlier breakdown of
  the system (dotted vertical line) due to $\jt_{23}$ dropping to zero (c),
  leading to divergent tunneling rates (e) and on-site energies (f). Note that
  Figs.~(a) and (d) have two separate vertical axes due to the different ranges
  of their observables.}%
  \label{fig:N110}%
\end{figure*}

In Fig.~\ref{fig:N110} the results of the many-body conditions \eqref{seq:mu_MB}
are shown in comparison with the \ac{MF} limit discussed in
Sec.~\ref{subsec:realization/embedding} for a weakly interacting \ac{BEC}
described by the \ac{BH} model. The parameter $\gamma = 0.5$ is chosen in such a
manner that the time scale \eqref{eq:time_scale} for the \ac{MF} calculations
reduces to the ratio of the initial occupations of the left reservoir and
embedded system wells, $\tau = n_0/n_1$. In the \ac{MF} (dashed lines) we
clearly find stationary states in the embedded wells, which are in agreement
with \cite{Kreibich2013,Kreibich2014} and the analytical solutions from
Eq.~\eqref{seq:embedded_init} (see especially Figs.~\subref{fig:N110}{a--b}).
The occupations in the reservoir wells change linearly in time as in
Eqs.~\eqref{seq:stat_n_MF} and the parameters in Figs.~\subref{fig:N110}{e--f}
follow Eqs.~\eqref{seq:parameters_MF}. At $t = 10$ the calculation breaks down
due to the left system well being empty and $J_{01}$ thus diverges according to
Eq.~\eqref{eq:J01_MF}.

In the \ac{MF}, $\PT$ symmetry in the inner wells is realized by
quasi-stationary states for the whole time interval $t \le 10$; applying the
same conditions to the many-body system yields a somewhat shorter time scale due
to an earlier breakdown of the system at $t \sim 8$ without any well being
completely empty. This breakdown is caused by the reduced current $\jt_{23}$
dropping to zero, which is a singularity in both $J_{23}$ and $\mu_3$ (see
Eqs.~\eqref{eq:J23_MF} and \eqref{eq:mu3_MB}). For $t \lesssim 2$, however, the
many-body dynamics is in good agreement with the \ac{MF} limit, showing
approximately the same stationary behavior. This is reasonable, since we always
start calculations with a pure state. However, eventually the dynamics of the
many-body system evolves differently due to decoherence, thus not realizing
quasi-stationary states anymore, as is clearly observable in
Figs.~\subref{fig:N110}{a--d}.

However, the differences between \ac{MF} and many-body dynamics as well as the
time of the breakdown however strongly depend on the choice of parameters, in
particular the total number of particles $N$ and the contact interaction
strength $g$. If the ratio $g/N$ becomes small, the influence of the
many-particle contributions becomes insignificant and the \ac{MF} description is
good. The larger this ratio becomes, the earlier the many-particle dynamics will
deviate from the stationary solutions and the system will break down.

It should be noted, that Eqs.~\eqref{eq:c_eq_0} hold for all times (see
Fig.~\subref{fig:N110}{d}), since we demanded it reference to the conditions in
the \ac{MF}. Thus again, these conditions are only physically meaningful for
balanced gain and loss in the vicinity of pure states and do not prevent the
system from decoherence. Instead, Eqs.~\eqref{eq:c_eq_0} are maintained by a
non-stationary many-body dynamics.

\subsection{Realization of balanced gain and loss}
\label{subsec:results/BGL}

To achieve true quasi-stationarity in the many-body system for $t \le \tau$,
which corresponds to exact $\PT$ symmetry in the related \ac{MF} system, we have
to demand that $\pdiff{}{t}\sigma_{12} = 0$, i.e.
\begin{subequations}
\begin{align}
  \rpart\parens*{\pdiff{}{t}\sigma_{12}} &= \frac{1}{2} \pdiff{}{t} \cx{1}_{12}
    = 0 \comma \label{eq:stat_c12} \\
  \ipart\parens*{\pdiff{}{t}\sigma_{12}} &= \frac{1}{2} \pdiff{}{t} \jtx{1}_{12}
    = 0 \period \label{eq:stat_jt12}
\end{align}
\label{seq:stat_sigma12}%
\end{subequations}
Since $J_{12}$ is time-independent, Eq.~\eqref{eq:stat_jt12} states that the
current between the inner wells is constant, directly implicating stationary
occupation numbers in said wells. In doing so one has to keep in mind that in
case of a many-body system stationarity can only refer to the expectation
values, since the variances only vanish completely in the \ac{MF} limit. In
fact, variances are closely connected to the higher-order terms in the BBGKY
hierarchy,
\begin{subequations}
\begin{align}
  \var\parens{\operator{n}_i} &= \cov_{iiii} \label{eq:var_n} \comma \\
  \var\parens[\big]{\ojt_{ij}} &= \cov_{ijji} + \cov_{jiij} - \cov_{ijij}
    - \cov_{jiji} \comma \label{eq:var_jt} \\
  \var\parens[\big]{\operator{c}_{ij}} &= \cov_{ijji} + \cov_{jiij} + \cov_{ijij}
    + \cov_{jiji} \comma \label{eq:var_c}
\end{align}
where
\begin{equation}
  \cov_{ijkl} = \tpdm_{ijkl} - \spdm_{ij} \spdm_{kl}
  \label{eq:covariance}
\end{equation}
\label{seq:variances}%
\end{subequations}
are the covariances \cite{Dast2014,Dast2016a,Dast2016b}.

\begin{figure}
  \includegraphics[width=\columnwidth]{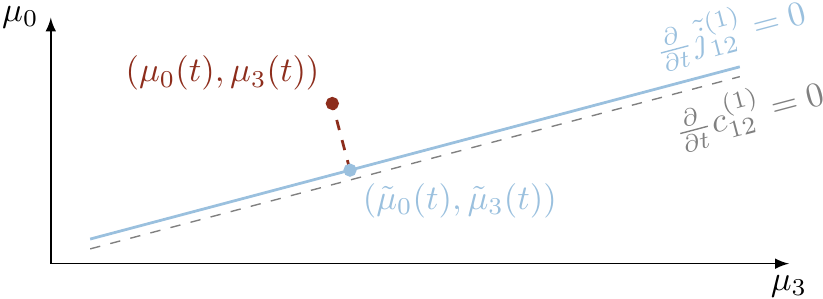}
  \caption{Eqs.~\eqref{seq:stat_sigma12} in the vicinity of pure states. Since
  both lines are almost parallel, the unique solution cannot be calculated
  accurately. By projection of the parameters \eqref{seq:mu_MB} onto the
  imaginary part yielding $(\tilde{\mu}_0,\tilde{\mu}_3)$ a constant particle
  flow between the inner wells is achieved. While in principle all points on the
  solid curve representing Eq.~\eqref{eq:stat_jt12} are valid, our approach
  stays as close as possible to the former approach adapted from the \ac{MF}.}
  \label{fig:modification}
\end{figure}

As one may easily confirm by using Eq.~\eqref{eq:TMM}, both
Eqs.~\eqref{seq:stat_sigma12} are fulfilled in the \ac{MF} limit at the same
time by $n_1 = n_2$, i.e., they are dependent in the \ac{MF}. In contrast, these
equations are independent in the many-body case. This makes these conditions
suitable to determine the on-site energies $\mu_0$ and $\mu_3$. Although this
method should thus provide unique solutions for the parameters, the fact,
however, that both conditions coincide for pure states renders any numerical
calculation impossible. Eqs.~\eqref{seq:stat_sigma12} are linear equations in
the parameter space $(\mu_0,\mu_3)$ (cf.\ Eqs.~\eqref{seq:BBGKY}), therefore the
unique solution is given by the intersection. In the vicinity of pure states,
however, those lines are approximately parallel as outlined in
Fig.~\ref{fig:modification}. In this case the ability to accurately calculate
the intersection point is lost due to numerical errors. Using these parameters
in calculations will eventually destroy the dynamics after only a few
iterations.

Obviously our approach does not allow for a realization of full
quasi-stationarity in a many-body system with pure initial states, i.e.,
stationary expectation values in the embedded system, which again corresponds to
$\PT$-symmetry in the \ac{MF} limit; note, however, that this method should work
for sufficiently impure states. Instead, we want to pursue the realization of
\ac{BGL} in the sense that there is a stationary, non-zero current expectation
value between the inner wells (see Eq.~\eqref{eq:stat_jt12}). This again leads
to the conservation of the occupation expectation values in the embedded wells
for a finite time, provided that the system is initially prepared in a state of
the form \eqref{seq:init_MF}. To achieve this, the on-site energies are modified
according to Fig.~\ref{fig:modification}, choosing a specific solution with
stationary current which is closest to the parameters determined via
Eqs.~\eqref{seq:mu_MB}. The result is
\begin{subequations}
\begin{align}
  \mu_0^{(1)} &= \frac{\alpha \Omega + \beta^2 \mu_0^{(0)}
    - \alpha \beta \mu_3^{(0)}}{\alpha^2 + \beta^2} \comma
    \label{eq:mu0_modified} \\
  \mu_3^{(1)} &= \frac{\beta \Omega + \alpha^2 \mu_3^{(0)}
    - \alpha \beta \mu_0^{(0)}}{\alpha^2 + \beta^2} \comma
    \label{eq:mu3_modified}
\end{align}
where
\begin{align}
  \alpha &= J_{01} \parens*{\jtx{1}_{02}
    + \cx{1}_{02} \frac{\cx{1}_{01}}{\jtx{1}_{01}}} \comma \label{eq:alpha} \\
  \beta &= J_{23} \parens*{\jtx{1}_{13}
    + \cx{1}_{13} \frac{\cx{1}_{23}}{\jtx{1}_{23}}} \comma \label{eq:beta} \\
  \Omega &= J_{01} \cx{1}_{02} \frac{\Chi{1}_{01}}{\jtx{1}_{01}}
    - J_{23} \cx{1}_{13} \frac{\Chi{1}_{23}}{\jtx{1}_{23}} \nonumber\\
  &\quad + J_{01} \Ups{1}_{02} + J_{12} \parens*{\Ups{1}_{22}
    - \Ups{1}_{11}} - J_{23} \Ups{1}_{13} \nonumber\\
  &\quad - U \parens*{\Ups{2}_{1112} - \Ups{2}_{1222}} \label{eq:Omega}
\end{align}
and
\begin{equation}
  \Chi{1}_{ij} = 2\rpart{\Zeta{1}_{ij}}
  \label{eq:Chi}
\end{equation}
\label{seq:mu_mod}%
\end{subequations}
is defined in analogy to Eq.~\eqref{eq:Ups}.

Since Eq.~\eqref{eq:stat_c12} will not be fulfilled in general, i.e.,
$\cx{1}_{12}$ is not stationary, this method does not reproduce quasi-stationary
states reliably.

\subsubsection{Low particle numbers}

\begin{figure*}
  \includegraphics[width=0.9\textwidth]{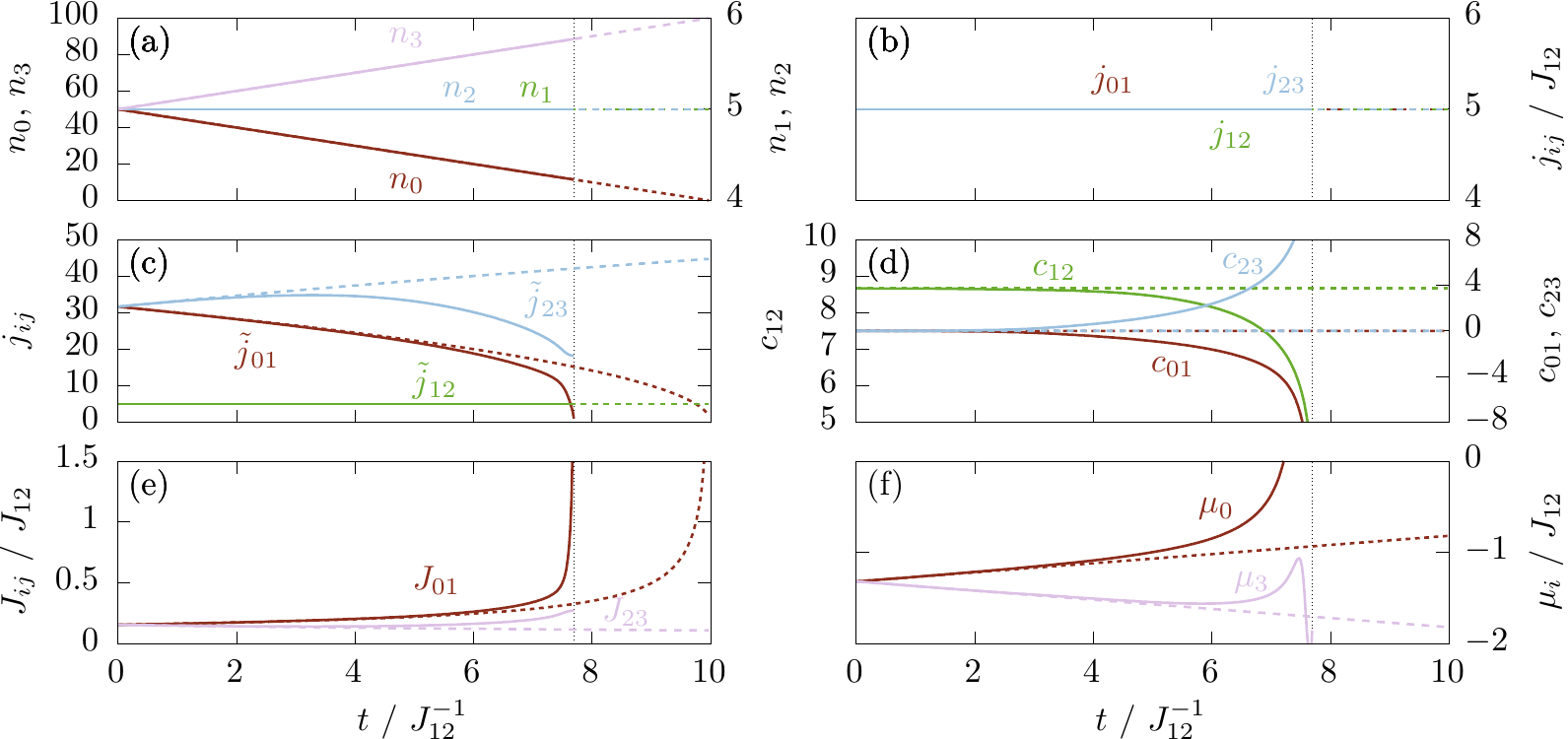}%
  \caption{Same as Fig.~\ref{fig:N110} but with modified on-site energies as
  shown in Fig.~\ref{fig:modification}. The \ac{BH} dynamics (solid lines) show
  stationary occupation numbers (a) in the embedded wells and stationary
  currents (b) between neighboring wells, which we call the \ac{BGL} state.
  There is still an earlier breakdown in comparison with the \ac{MF} dynamics
  (dashed lines), now caused by $\jt_{01}$ dropping to zero. Due to the
  modifications \eqref{seq:mu_mod}, some parameters are no longer monotonous
  (f). Note that Figs.~(a) and (d) have two separate vertical axes due to the
  different ranges of their observables.}%
  \label{fig:N110_statcurr}%
\end{figure*}

The dynamics obtained with these modifications for a relatively low total number
of $N = 110$ particles is shown in Fig.~\ref{fig:N110_statcurr} with the same
initial conditions as in Fig.~\ref{fig:N110} and also in comparison with the
\ac{MF} dynamics. The many-body system again follows the \ac{MF} dynamics on a
small time scale $t \lesssim 2$ almost exactly and then deviates leading also in
this case to an earlier breakdown than in the \ac{MF}. However, for the whole
time scale before the breakdown of the system both the occupation and current
expectation values, i.e., $n_1$, $n_2$ and $\jt_{12}$, are preserved exactly,
which can clearly be seen in Figs.~\subref{fig:N110_statcurr}{a--b}. It is
remarkable that not only the quantities in the embedded system are in perfect
agreement with the \ac{MF} dynamics, but that the same also holds for all other
occupation and current expectation values; this means, the occupation numbers in
the reservoir change linearly in time and all current expectation values are
equal and conserved, cf.\ Eqs.~\eqref{seq:embedded_init} and
\eqref{seq:stat_n_MF}.

Apart from this, since $c_{12}(t)$ is explicitly time-dependent and thus in
general changes in time, the realized state is not truly quasi-stationary. That
is, in a many-body system \ac{BGL}, as we introduced it, will not induce
quasi-stationarity by default anymore as it is the case in the \ac{MF}.
Nevertheless, these states will make a transition into true quasi-stationary
states in the \ac{MF} limit, i.e., $\PT$-symmetric states of the embedded
system. It should be noted that, due to the modified on-site energies
\eqref{seq:mu_mod}, the parameters may show more complex behavior, as can be
seen for example in Fig.~\subref{fig:N110_statcurr}{f}. Besides, the breakdown
is, unlike before, now caused by the reduced current $\jt_{01}$ dropping to
zero, as can be seen in Fig.~\subref{fig:N110_statcurr}{c}, leading to
singularities in $J_{01}$ and $\mu_0$ (see Eqs.~\eqref{eq:J01_MF} and
\eqref{eq:mu0_MB}).

\subsubsection{Large particle numbers}

As discussed in Sec.~\ref{subsec:MB_models/BBR}, the computational effort for
calculating exact \ac{BH} dynamics increases rapidly with the total number of
particles restricting us to $N \lesssim 200$. To investigate systems with a
total number of particles on the same order of magnitude as in typical
experiments, i.e., $N \gtrsim 1000$, we will use the \ac{BBR} method introduced
in Sec.~\ref{subsec:MB_models/BBR}. It should be noted at this point, that the
\ac{BBR} approximation for $N \sim 200$ already shows excellent agreement with
exact many-body calculations on the given time scale, therefore justifying its
application for larger particle numbers.

\begin{figure*}
  \includegraphics[width=0.9\textwidth]{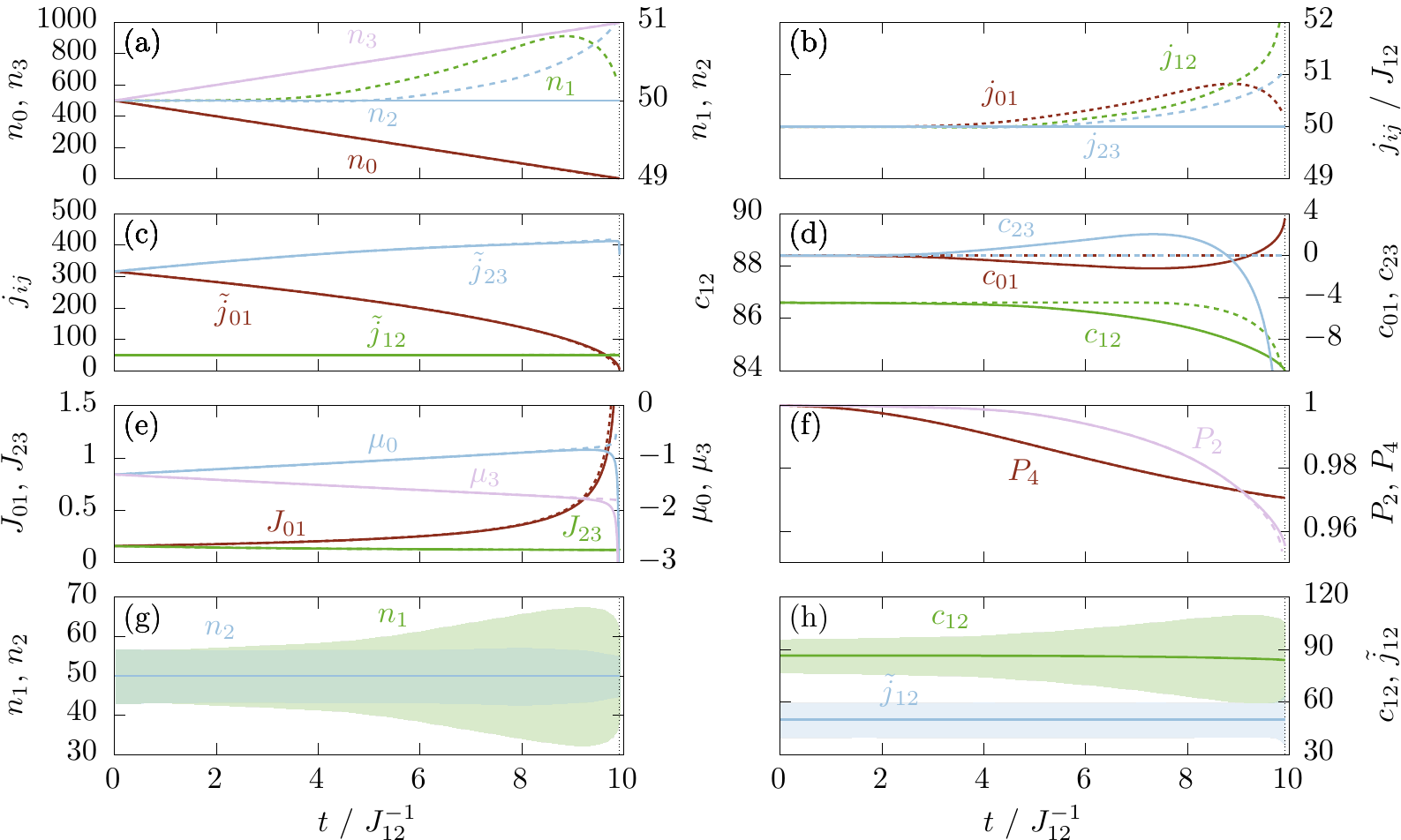}%
  \caption{Comparison of the first approach from Fig.~\ref{fig:N110} (dashed
  lines) with the second approach from Fig.~\ref{fig:N110_statcurr} (solid
  lines) for a total number of $1100$ particles with the same occupation ratio
  as before. Many-body calculations were performed within the \ac{BBR}
  approximation. Although the time of the breakdown (dotted vertical line)
  almost matches that of the \ac{MF} limit in Figs.~\ref{fig:N110} and
  \ref{fig:N110_statcurr} (note that the calculation ends slightly before $t =
  10$), there are still deviations from the \ac{MF} dynamics (a--d), cf.
  Figs.~\ref{fig:N110} and \ref{fig:N110_statcurr}. For both approaches the
  time-dependent parameters are in good agreement and only show different
  behavior close to the breakdown. The same holds for the purities of the
  four-mode system $P_4$ and the embedded two-mode system $P_2$ (f). The
  standard deviations for the embedded observables are shown for the \ac{BGL}
  state (g--h). Note that Figs.~(a), (d), and (e) have two separate vertical
  axes due to the different ranges of their observables.}%
  \label{fig:N1100}%
\end{figure*}

Fig.~\ref{fig:N1100} shows a comparison between the two approaches from
Figs.~\ref{fig:N110} and \ref{fig:N110_statcurr}, but for $N = 1100$. For such
large numbers of particles the relative deviations between the different
many-body calculations become very small and the same holds for deviations from
the \ac{MF} dynamics. Nonetheless, the key differences are still observable,
i.e., the system moves away from the \ac{BGL} state after only few time steps
without modification of the on-site energies. While these differences are quite
prominent in the observables shown in Figs.~\subref{fig:N1100}{a--d}, i.e., the
system evolves in a very distinctive manner in each case, it is rather
surprising that there are only minor differences in the parameters shown in
Fig.~\subref{fig:N1100}{e} for the most part of the time evolution. Only towards
the breakdown of the system the behavior changes due to the divergent behavior,
the modifications thus becoming more significant.

Using the reduced \ac{SPDM} $\spdm_\mathrm{red}$ and the fact that it has only
one macroscopic eigenvalue in case of a pure \ac{BEC}, corresponding to a
general criterion for Bose-Einstein condensation
\cite{Penrose1951,Landau1951,Penrose1956,Yang1962}, we can define the purity
\cite{Dast2016a} of an $M$-mode system by
\begin{equation}
  P_M = \frac{M \tr\parens*{\spdm_\mathrm{red} \cdot \spdm_\mathrm{red}} - 1}{M - 1} \period
  \label{eq:purity}
\end{equation}
This quantity is restricted to $\brackets{0,1}$, where $P_M = 1$ describes a
perfectly pure \ac{BEC} and $P_M = 0$ a maximally fractioned and thus impure
\ac{BEC}. Fig.~\subref{fig:N1100}{f} shows the purities of the four-mode system
$P_4$ and of the embedded two-mode system $P_2$ under the assumption that the
embedded system effectively models an open quantum system. Also in this case the
deviations between the original and the modified method remain small. Due to
particle interactions, the coherence of the initial state is eventually
destroyed \cite{Ruostekoski1998}, i.e., the purity of the system is reduced. On
the one hand, it is remarkable that the purity of the embedded system drops only
slightly at first despite the rather small number of $N_2 = 100$ particles,
which is a consequence of the initial quasi-stationary state of the
corresponding \ac{MF} system \cite{Dast2014,Dast2016a}. On the other hand, this
shows that \ac{BGL} can be realized in the true many-body regime, in which most
part of the simulation takes place.

Figs.~\subref{fig:N1100}{g--h} show the observables of the embedded system for
the modified on-site energies in the context of their respective variances
\eqref{seq:variances}. The variances are on the same order of magnitude as the
total particle number $N$, even though, for increasing $N$ the relative
variances of the observables will shrink. More interesting, however, is the
different behavior of the occupation number expectation values. The expectation
values of $n_1$ and $n_2$ are conserved for the whole simulation time, and their
variances are initially equal. Yet, we then find an overall growth of
$\var\parens{n_1}$ while $\var\parens{n_2}$ remains approximately constant.
Similar behavior is also visible for other one-particle expectation values such
as the reduced currents between system and reservoir \cite{Dizdarevic2016}.
While such a growth of variances over time, that is, an increasing uncertainty,
may be a consequence of the finite reservoir and the constraints due to the
time-dependent parameters, it certainly is connected to the breakdown of the
system caused by a singularity due to the vanishing of a reduced current
expectation value.

However, even if time-dependent optical lattices are theoretically feasible and
experimentally accessible for the control and realization of such systems, an
experimental realization would be fundamentally different from the theoretical
methods. In real experiments one would probably tune the parameters in a
predefined way, such that the potential is only time- but not state-dependent.
That is, one has to determine the parameters by theory rather than by means of
\textit{in-situ} measurements, which are collapsing the wave function and thus
destroying the condensate immediately. In this way, the system will evolve
unperturbed and can be measured at different times, so that comparing the
observables to theoretical results in the limit of many realizations of the
experiment is possible.

\section{Conclusions}
\label{sec:conclusions}

By means of time-dependent potentials in four-mode optical lattices, we
investigated the possibilities of effectively describing pure two-mode \acp{BEC}
with gain and loss in the $\PT$-symmetric regime. Starting with the \ac{GPE} for
pure states in the \ac{MF} limit we derived precise conditions leading to
time-dependent parameters. To further investigate the few-body limit, we used
exact time evolution to perform many-body calculations of our system. These
simulations, however, show that the \ac{MF} approach is not suitable in the
many-body case and the system evolves into the $\PT$-broken regime albeit being
in the $\PT$ symmetric regime at the beginning. As discussed in
Sec.~\ref{subsec:results/BGL}, we cannot achieve true quasi-stationarity by
means of pure initial states.

Nevertheless we realized what is probably the intuitive notion of \ac{BGL},
i.e., a configuration where stationary populations and a constant particle flow
are present for the whole simulation time. Yet, also in this case the system
will break down at some point into a $\PT$-broken state. The creation of
controlled gain and loss over such long timescales lays the foundation for
experimental investigations of generic many-body quantum effects in \ac{BGL}
systems like the aforementioned purity oscillations
\cite{Dast2016a,Dast2016b,Dast2017}.

We can further confirm with our system that exact non-oscillating
$\PT$-symmetric pure states, which arise very naturally in the \ac{MF} limit as
the unique stationary solutions in an open quantum system, are an exclusive
feature of the \ac{GPE}, as previous many-body investigations of the two-mode
model have already shown \cite{Dast2014,Dast2017}. Although no true
quasi-stationarity can be achieved for finite numbers of particles, that is
without the \ac{MF} approximation, both approaches discussed in
Sec.~\ref{sec:results} will make a transition into the $\PT$-symmetric four-mode
model in the \ac{MF} limit.

In the present work we used the minimum setup for the realization of reservoirs,
i.e., only one reservoir well at each side, respectively. Although such a
reservoir may behave approximately Markovian for a large number of particles,
which can be investigated using the \ac{BBR} approximation, we still see
prominent differences. Since every additional reservoir well is increasingly
expensive in computational costs, a large number of reservoir wells cannot be
treated without further approximation. In a future work these additional
reservoir wells could be included into our many-body calculation by means of the
\ac{MF} limit. In this way, effects of the finite reservoir could be suppressed.
Another improvement might be the inclusion of a real loss term as used in
\cite{Barontini2013,Santra2015,Labouvie2016}, rendering a reservoir at the
out-coupling well unnecessary.

%%%%%%%%%%%%%%%%%%%%%%%%%%%%%%%%%%%%%%%%%%%%%%%%%%%%%%%%%%%%%%%%%%%%%%%%%%%%%%%%

% Specify following sections are appendices. Use \appendix* if there
% only one appendix.
%\appendix
%\section{}

% If you have acknowledgments, this puts in the proper section head.
%\begin{acknowledgments}
% put your acknowledgments here.
%\end{acknowledgments}

\appendix*
\section{Lexicographic hopping}
\label{app:lexicographic_hopping}

To calculate \ac{BH} dynamics, the Hamiltonian has to be applied directly by
matrix multiplication, which means, the entries of the Hamiltonian matrix have
to be calculated. While the potential part of the Hamiltonian \eqref{eq:BHM}
yields only diagonal elements, the kinetic terms are off-diagonal. Since the
resulting matrix is extremely sparse, only the non-zero entries have to be
calculated explicitly. This implies that one has to apply every
creation-annihilation operator pair, i.e., the hopping terms, to each Fock state
separately and find the index of the resulting state. Searching for specific
Fock states in a large basis set is slowing down calculations considerably. We
will therefore present a method for efficiently calculating the state index in a
lexicographically ordered Fock basis. Details about this method and its
implementation can be found in Ref.\ \cite{Alpin2016}.

The dimension of the entire Fock basis for $N$ particles in $M$ potential wells is given by
\begin{equation}
  D(N, M) = \begin{pmatrix} N+M-1 \\ N \end{pmatrix} \period
  \label{eq:dimension}
\end{equation}
All Fock states with $n_1 < N$ particles on the first site are part of a
subspace with $\tilde{n} = N - n_1$ particles in $M - 1$ sites. These
$D(\tilde{n}, M')$ states start at index
\begin{equation}
  \nu(n_1) = \sum_{\tilde{n} = 0}^{\mathclap{N-n_1-1}} D(\tilde{n}, M-1)
  \label{eq:index_1}
\end{equation}
such that the first Fock state with $n_1 = N$ occupies $\nu = 0$. For a Fock
state $\ket{n_1, \ldots, n_M}$ we have to sum up over all sites, which
eventually yields the global index
\begin{equation}
  \nu(n_1, \ldots, n_M) = \sum_{m = 1}^{M - 1} \sum_{\tilde{n} = 0}^{N_m - 1}
    D\parens*{\tilde{n}, M_m} \comma
  \label{eq:lexikographic_index}
\end{equation}
where $N_m = N-\sum_{k = 1}^m n_k$ and $M_m = M - m$. It should be noted, that
only the first $M-1$ occupation numbers are important, since the last one is
included in the boundary condition $\sum_{m = 1}^M n_m = N$.

The numerical effort can further be reduced, when only the index shift between
the Fock states before and after the application of a hopping term $\ad_i\a_j$
is calculated, which is given by
\begin{equation}
  s_{ij} = \begin{cases}
              \displaystyle - \sum_{m = \tilde{m}}^{\tilde{M}-1}
                D\parens*{N_m-1, M_m} & n'_{\tilde{m}} > n_{\tilde{m}} \\
              \displaystyle \phantom{-} \sum_{m = \tilde{m}}^{\tilde{M}-1}
                D\parens*{N_m, M_m} & n'_{\tilde{m}} < n_{\tilde{m}}
            \end{cases}
  \label{eq:LJS}
\end{equation}
with $\tilde{m} \equiv \min\parens{i, j}$ and $\tilde{M} \equiv \max\parens{i,
j}$. Since the kinetic part of the \ac{BH} model \eqref{eq:BHM} only contains
nearest-neighbor hops, Eq.~\eqref{eq:LJS} reduces to the calculation of a single
binomial coefficient. Non-nearest-neighbor hops on the other hand can be
interpreted as a sequence of successive nearest-neighbor hops.

% Create the reference section using BibTeX:
% ****** Start of file paper.bbl ****** %
%merlin.mbs apsrev4-1.bst 2010-07-25 4.21a (PWD, AO, DPC) hacked
%Control: key (0)
%Control: author (72) initials jnrlst
%Control: editor formatted (1) identically to author
%Control: production of article title (-1) disabled
%Control: page (0) single
%Control: year (1) truncated
%Control: production of eprint (0) enabled
%
% ****** End of file paper.bbl ****** %

\end{document}